\documentclass{elsart}

\usepackage{amssymb}
\usepackage{epsf,colordvi,color,amsbsy}
\usepackage[dvips]{graphicx}
\begin{document}

\begin{frontmatter}

\title{Reconstructed Total Solar Irradiance as a precursor for long-term solar activity predictions: a nonlinear dynamics approach}

\author[]{Stefano Sello \corauthref{}}

\corauth[]{stefano.sello@enel.it}

\address{Mathematical and Physical Models, Enel Research, Pisa - Italy}

\begin{abstract}
Total solar irradiance variations, about $0.1\%$ between solar activity maximum and minimum, are available from accurate satellite measurements
since 1978 and thus do not provide useful information on longer-term secular trends. Recently, Krivova et al., 2007 reconstructed, using suitable models, the total solar irradiance from the end of the Maunder minimum
to the present, based on variations of the surface distribution of the solar magnetic field. The latter is calculated from the long historical record of the sunspot numbers using a simple
but consistent physical model.
There are many classes of proposed prediction methods for solar cycles behavior, based on different direct solar activity indices or on various valuable proxies. In particular, the precursor based methods, utilize a given proxy index to predict the future evolution
of solar activity. Long-term time series of sunspot numbers, allow us to reliably predict the behavior of the next solar cycle, few years in advance.
In previous papers we predicted the full-shape curve of future solar cycles, using a proper non-linear dynamics method applied to monthly smoothed sunspot numbers.
In particular, we proved that a sufficiently reliable phase-amplitude prediction of the current solar cycle 24, requires the knowledge of the initial increasing phase of the cycle spanning at least two years.
The aim of the present paper is to give a robust long-term prediction of solar cycle activity, many years in advance and for at least three successive solar cycles using the same nonlinear method and, as solar activity proxy index, the
reconstructed total solar irrandiance.

\end{abstract}
\end{frontmatter}

\section{Introduction}

Solar cycle full-shape prediction methods aim to determine the approximate whole cycle curve, not only its peak magnitude and timing. This task is particularly useful for numerous
scientific and technological applications. The main involved areas are the electric power transmission systems, airline and satellite communications, GPS signals, and extra-vehicular-activities of astronauts during space missions and, more in general, all the solar-terrestrial interactions.
On the other hand, it is well known the difficulties in predicting the full evolution of future solar cycles, due to highly complex non-linear dynamical processes involved, mainly related to interaction of different components, deterministic and stochastic ones, of the internal magnetic fields.
The predictions of the "anomalous" current solar cycle 24 allow us to deeply test our current knowledge on the complex dynamics of solar activity evolution and to further refine both our current solar dynamo models and new mathematical/numerical methods.
The main detected anomaly of current cycle 24 is related to its extended period of deep minimum solar activity, with very few small and weak sunspots and flares. There is now a general consensus about a clear diminished level of solar activity, suggesting a current transition phase between a long period of maximum and a next long period of 
moderately or weak activity (see de Jager and Duhau, 2009). In the modern era there is no precedent for such a protracted activity minimum, but there are historical records from a century ago of a similar pattern. 
Based on the solar radio flux F10.7 cm activity index, there seems to be little question that the new-cycle activity 24, started late in late 2008 or in early 2009. 
The unusual character of current solar cycle is also documented by the wide range of the predicted values using different prediction methods as well documented in the literature.
A comprehensive collection of predictions for solar cycle 24 is described by Pesnell and regularly released by NOAA/SEC international panel (Pesnell, NOAA/SEC panel).
In a previous paper (Sello, 2012), we predicted the full-shape curve of future solar cycle, using a proper non-linear dynamics method applied to monthly smoothed sunspot numbers.
In particular, we proved that a sufficiently reliable phase-amplitude prediction of the current solar cycle 24, requires the knowledge of the initial increasing phase of the cycle, after its long minimum, spanning at least two years.
In this paper we propose a robust long-term prediction of solar cycle activity, many years in advance and for at least three successive solar cycles using the same nonlinear method and, as proxy index, the
reconstructed long-term total solar irrandiance.

\section{Reconstructed Total Solar Irradiance (TSI)}

The  so-called "solar constant" is, in fact, not constant. Recent satellite observations, since 1978, have found that the Total Solar Irradiance (TSI), 
i.e. the amount of spectrum integrated solar radiation received at the top of the Earth's atmosphere, does vary at a level of about $0.1\%$ from maximum and minimum phases of the
solar cycle. The TSI provides the energy that determines the Earth's climate. In order to use the TSI time series as a possible 
proxy index for solar cycle predictions, we need a reliable reconstruction of solar irradiance for the pre-satellite period with the help of models.
A number of long-term (up to $300-400$ years) reconstructions, both purely empirical and more physics-based, have been produced in the last two decades, see e.g., Solanki and Krivova 2004; Lockwood
2005. Solanki et al., 2000, 2002, presented a simple physical model that allows a reconstruction of the solar total and open magnetic flux from the sunspot number
More recently, Krivova et al., 2007 use this model to reconstruct the solar total irradiance variations (TSI) back to 1700 using the Zurich
sunspot number and to 1610 using the Group sunspot number. In their approach the authors follow the successful irradiance modelling by,
e.g., Wenzler et al., 2006, as closely as possible and extend it to longer times. The authors also constrain the reconstructions by requiring
them to reproduce the available observed time series of total solar irradiance as well as total and open magnetic flux (for details see Krivova et al., 2007).
In the present work we use the above long-term reconstructed TSI time series as a proxy index for the input of the nonlinear dynamics method 
to predict, many years in advance, the solar cycle activity for at least three successive solar cycles, starting from 2005. In fact, the reconstructed
TSI time series by Krivova et al., 2007 consists of unevenly spaced daily values spanning the period: $1611.6.1-2004.12.31$ (file: tsi-1611.txt from:
http://www.mps.mpg.de/projects/sun-climate/data.html). For the solar activity predictions we used here both the TSI monthly averaged values and the TSI monthly
smoothed averaged values, given by the equation:

\begin{equation}
    <TSI_n>(t)=1/11 [{\sum_{k=n-5}^{k=n+5} TSI_k + {1 \over 2} (TSI_{n+6}+ TSI_{n-6})}]
\end{equation}

where: $TSI_k$ is the mean value of TSI for the month $k$.

Figure 1 shows both the reconstructed TSI monthly averaged and monthly smoothed averaged values since 1611 derived from Krivova et al., 2007.

\begin{figure}[h!]
\resizebox{\hsize}{!}{\includegraphics{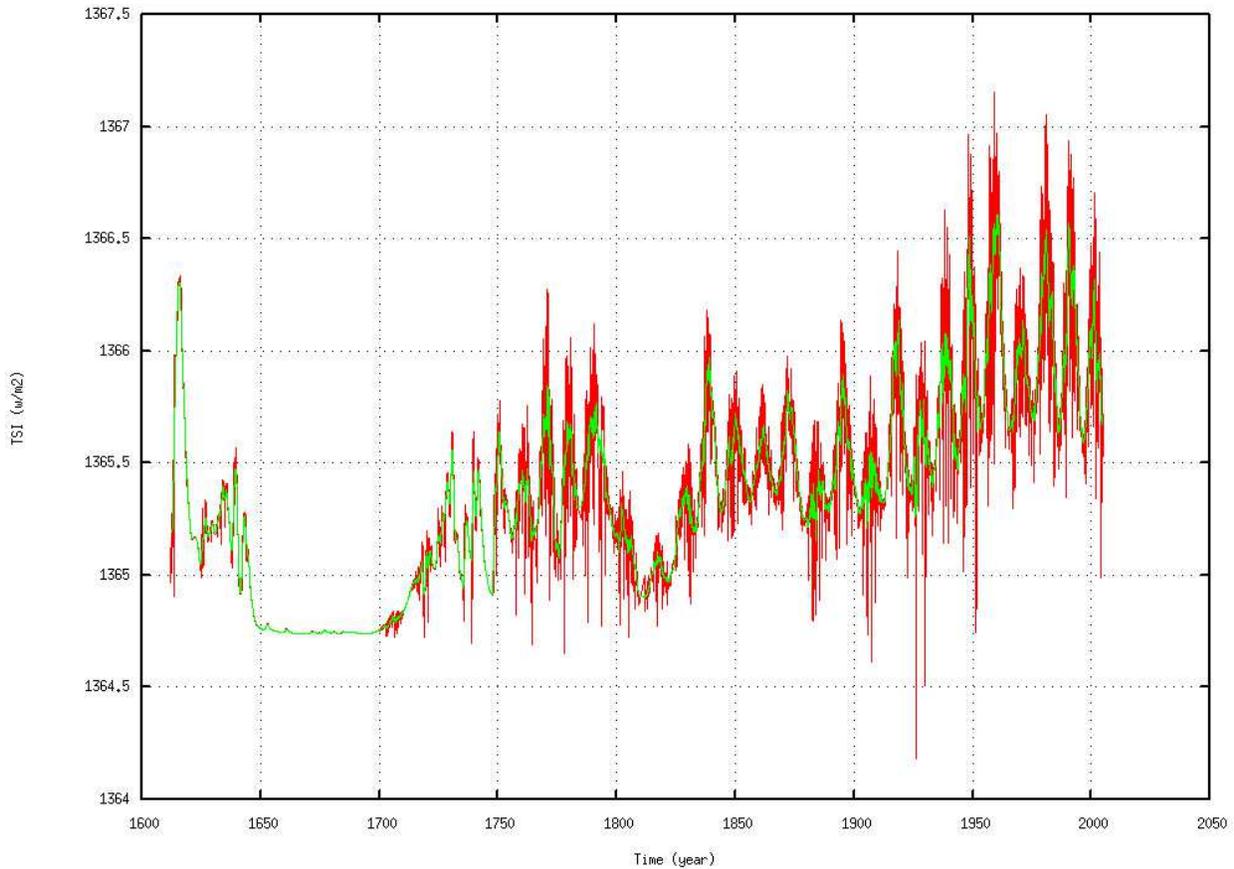}}
 \caption{The reconstructed TSI monthly averaged (red) and monthly smoothed averaged (green) values since 1611 (see Krivova et al., 2007).} 
 \label{fig1}
\end{figure}

\section{The radio flux at $10.7$ cm and its relation with TSI and sunspot numbers}

Since the reconstructed TSI time series is the proxy here used to predict the future solar activity, it is useful to consider the relation existing
between TSI and the solar radio flux F10.7 index and the more usual sunspot numbers.

The solar radio flux F10.7 index is a measure of the noise level generated by the sun at a wavelength of 10.7 cm (2800 MHz) at the earth's orbit.
The global daily value of this index is measured at local noon (1700 GMT) at the Ottawa/Pentictin Radio Observatory in Canada.
Historically, this index has been used as an input to ionospheric models as a surrogate for the solar output in wavelengths that produce
photoionization in the earth's ionosphere (in the ultraviolet bands). This radio flux, which originates from atmospheric layers high in the sun's chromosphere and low
in its corona, changes gradually from day-to-day, in response to the number of spot groups on the disk.  In fact, radio intensity levels consist of emission from three sources:  from the undisturbed solar surface, from developing active regions,
and from short-lived enhancements above the daily level. Solar flux density at 2800 megaHertz has been recorded routinely by radio telescope since
February 14, 1947. Here we used the set of absolute values i.e. without fluctuations from the changing sun-earth distance and reduced uncertainties in antenna gain and in waves reflected from the ground
(see NOAA Solar-Terrestrial Physics division, http://www.ngdc.noaa.gov/stp/stp.html).

Figure 2 shows both the monthly averaged and monthly smoothed F10.7 index behavior since 1947 from file: MONTHLY-ABS-F10.7.txt (NOAA Solar-Terrestrial Physics division). 

\begin{figure}[h!]
\resizebox{\hsize}{!}{\includegraphics{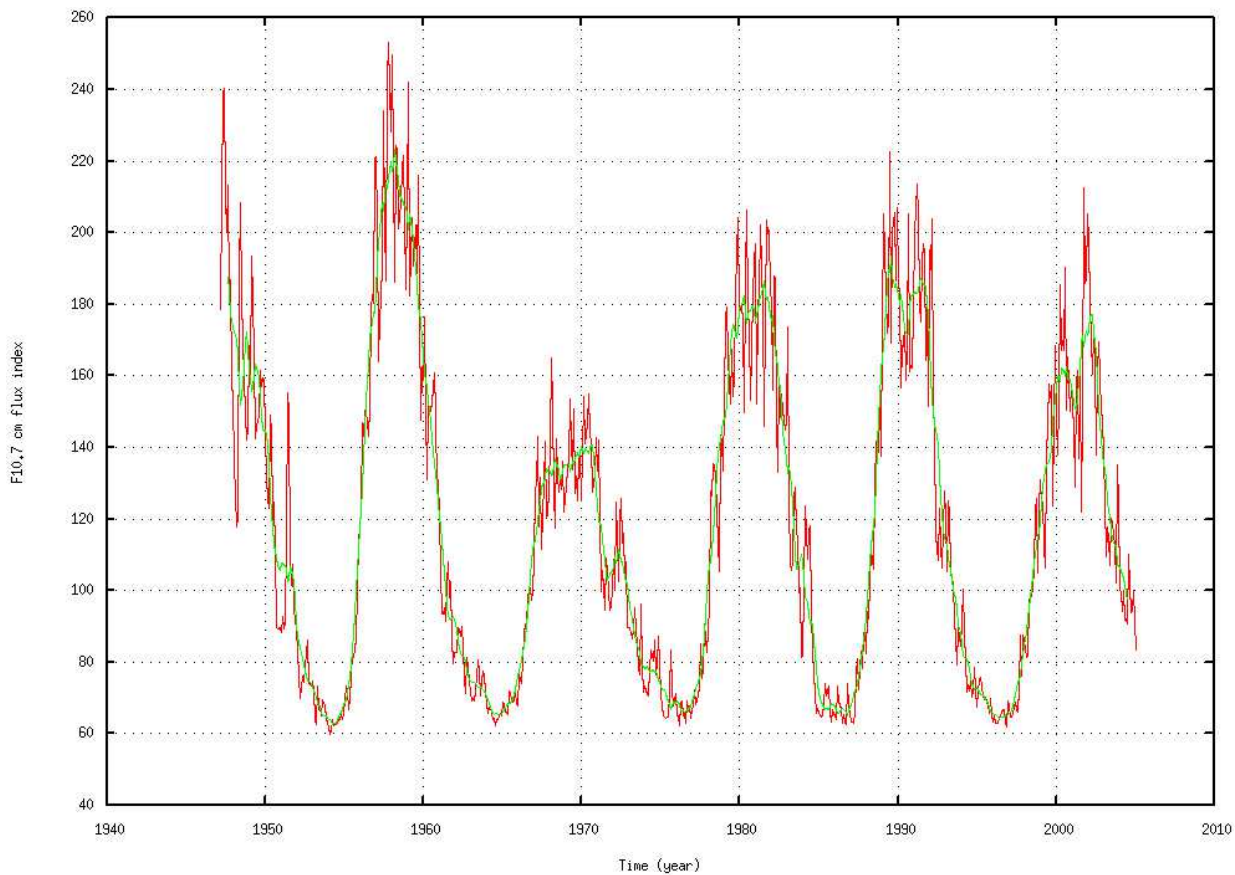}}
 \caption{The monthly averaged (red) and monthly smoothed averaged (green) F10.7 index values since 1947 from NOAA Solar-Terrestrial Physics division.}
 \label{fig2}
\end{figure}

In order to determine the relation between the TSI and F10.7 index, we computed the (linear) correlation existing between the successive maxima (peak values)
for both the smoothed averaged TSI and F10.7 index, since 1947, i.e. for about six solar cycles. Figure 3 shows the results of the correlation
analysis, with a resulting correlation coefficient: R=0.9. This relation will be useful in determining the link between the predicted TSI and the solar radio flux, a common
solar activity index.

\begin{figure}[h!]
\resizebox{\hsize}{!}{\includegraphics{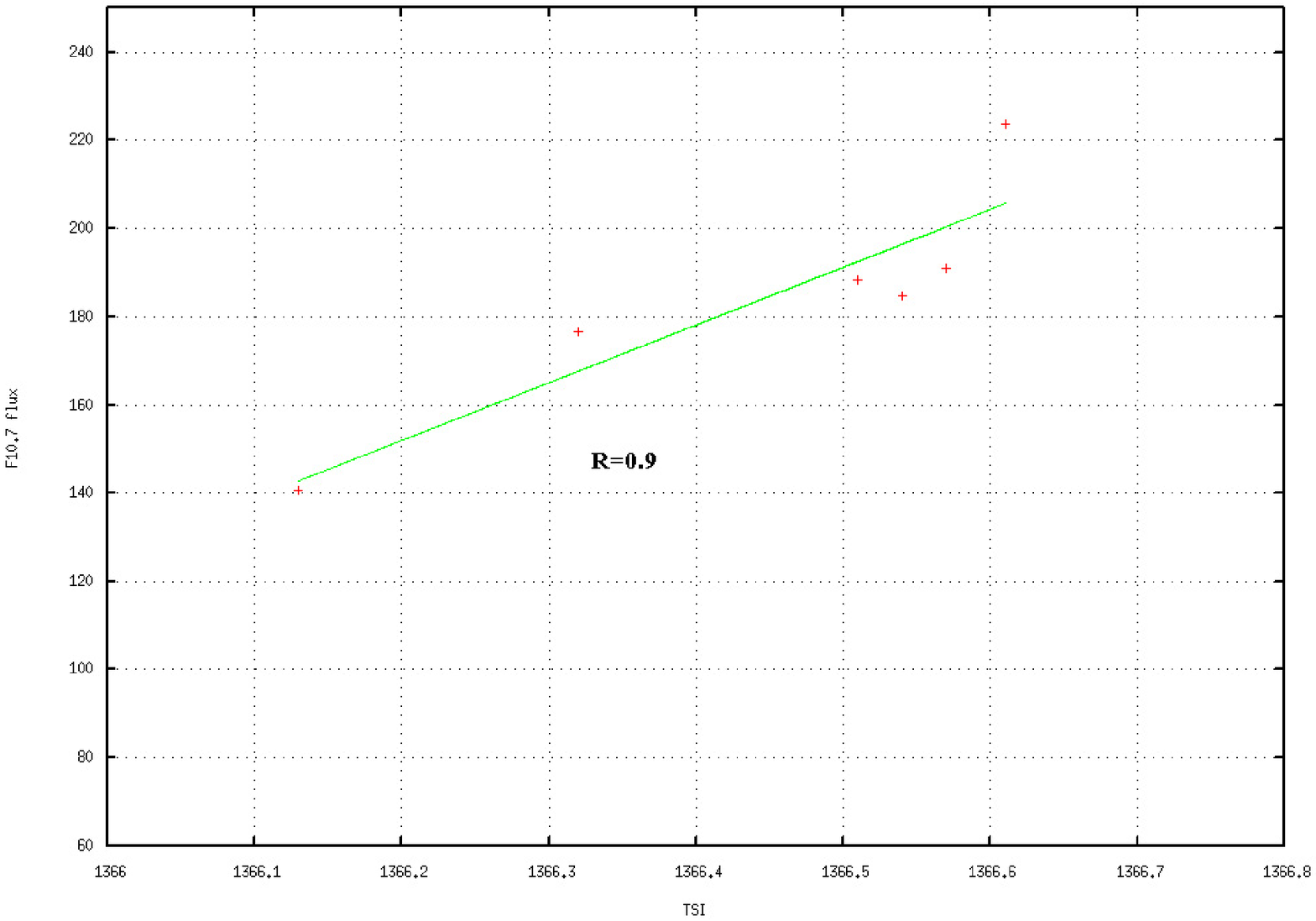}}
 \caption{The correlation existing between the successive maxima for both the smoothed averaged TSI and F10.7 index. The correlation coefficient
is: R=0.9 (green line).}
 \label{fig3}
\end{figure}

The next step is to relate the F10.7 index to the monthly smoothed sunspot number. It is well known that there is a strong (nonlinear) correlation between the F10.7 index and the sunspot numbers.
Figure 4 is a plot of the monthly-averaged sunspot number against the monthly-averaged F10.7 solar flux for data between 1947 and 1990.
The correlation between these quantities is evident but there is still considerable scatter even for monthly-averaged values. 
The following empirical equation is useful for converting the F10.7 flux (F) to sunspot number (R).

\begin{equation}
  R = 1.61 FD - (0.0733 FD)^2 + (0.0240 FD)^3 
\end{equation}

where: $FD = F - 67.0$. Of course, this equation is valid on a statistical (i.e. average) basis (see Thompson, IPS - Radio and Space Services, http://www.ips.gov.au/ ).

\begin{figure}[h!]
\resizebox{\hsize}{!}{\includegraphics{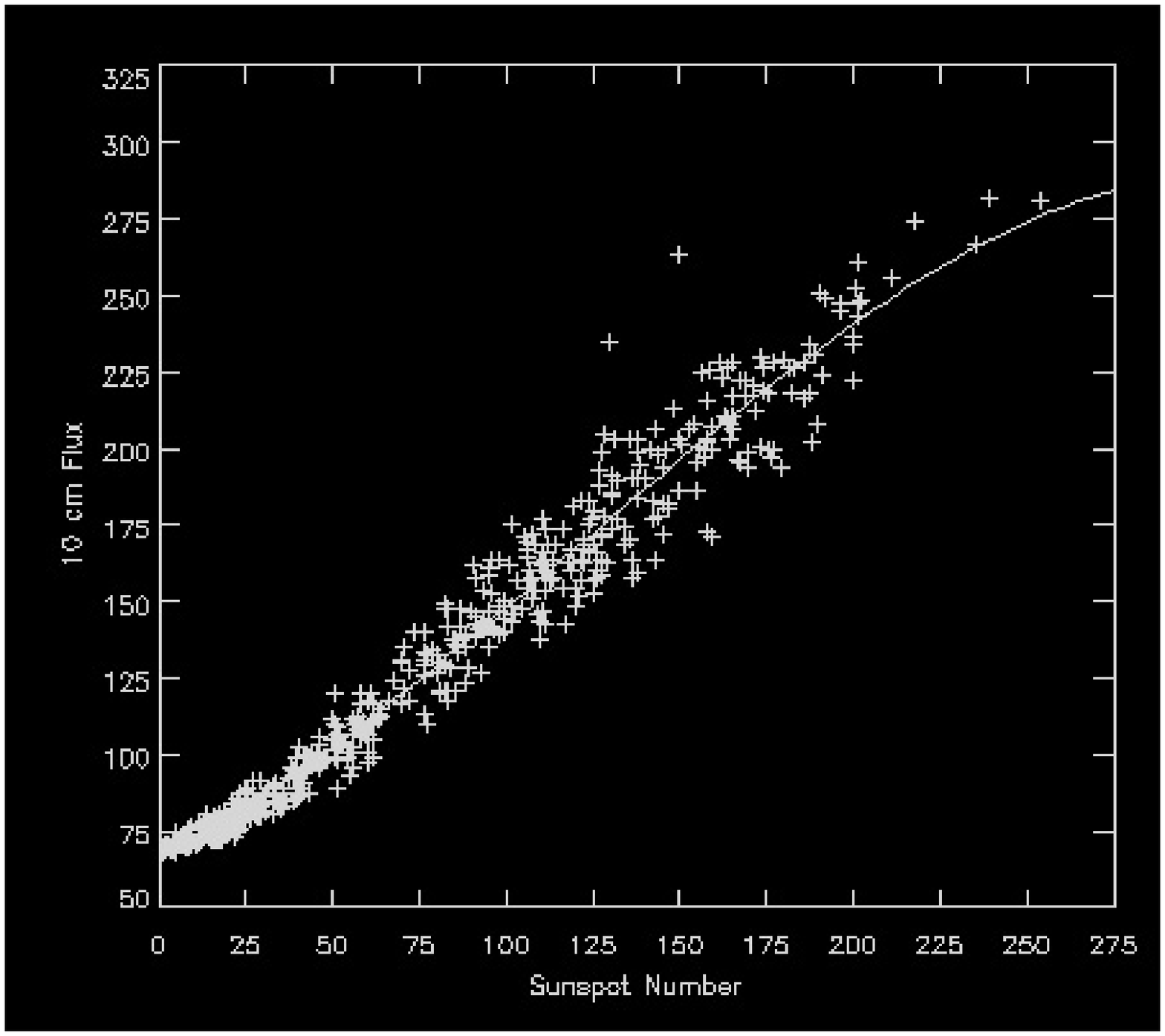}}
 \caption{The nonlinear correlation existing between the monthly averaged sunspot numbers and the solar radio flux at 10.7 cm (from IPS - Radio and Space Services).}
 \label{fig4}
\end{figure}

The above equation will be used to statistically relate the predicted TSI value and the monthly smoothed sunspot number.

\section{Nonlinear dynamics method for prediction of solar cycle activity}

The nonlinear features of the monthly mean sunspot numbers and, more generally, of the solar activity is now well established, as is widely documented
by many different works. In particular, many independent analyses showed strong evidence for low-dimensional deterministic
nonlinear-chaotic dynamics governing sunspot numbers at timescales: $T > 8$ years (Zhang 1994, 1995; Lawrence et al. 1995; Schreiber 1998). Furthermore, a genuine chaotic behaviour
for monthly averaged Wolf numbers, using a nonlinear forecasting method, has been shown by Rozelot (1995). Further, Kugiumtzis (1999) investigated
some properties of a standard and a refined surrogate technique by Prichard and Theiler (1994) to test the nonlinearity
in a real time series, showing that, for the annual sunspot numbers, there is strong evidence that nonlinear dynamics
are in fact present, enforcing also the idea that the sunspot numbers are to a first approximation proportional to the
squared magnetic field strength. More recently, Li and Sofia (2001), in order to explain the observed total solar irradiance
long-term modulation, proposed a semi-empirical model for the maximum magnetic field in the solar interior which nonlinearly relates the yearly mean sunspot
numbers. In a previous work, Sello, 2001, we used the method of surrogate data combined with the computation of linear and nonlinear
redundancies, in order to show that the monthly mean sunspot numbers contain true nonlinear dependencies (Palus 1995).
The nonlinearity analysis on the monthly mean sunspot numbers and more generally on the solar activity indices, clearly supports the use of a nonlinear
dynamic approach as possible prediction method. The usefulness and validity of this solar activity prediction approach has been well documented in previous works (sello, 2001, 2012).
Here we recall some basic elements necessary to apply the nonlinear prediction method to real time series, referring to the cited works for more details.

When a continuous scalar signal $x(t)$, here the reconstructed monthly averaged TSI, is measured at discrete time intervals, $t_s$, we obtain a single scalar time series:

\begin{equation}
 {x(t_0); ~ x(t_0 + t_s); ~ x(t_0 + 2t_s); ... ; ~ x(t_0 + Nt_s)}
  \end{equation}

Starting from this time series (if sufficiently long) it is possible to reconstruct a \emph{pseudo-trajectory} in an m-dimensional embedding space
through the vectors (embedding vectors):

\begin{equation}
 \textbf{$f(y_s)$}= ( x(t_0+(s-1)t_s), ~ x(t_0+(s-1)t_s + \tau), ..., ~ x(t_0+(s-1)t_s + (m-1) \tau) )^T
\end{equation}

where: $s = 1, 2, ...,(1-m) {\tau \over {t_s}} + (1+N)$, and $\tau$ is the delay time.

A selection of suitable values for these embedding parameters in the reconstruction procedure is very crucial for the reliability and accuracy 
of results, as was pointed out in many works (Shuster 1984; Theiler, 1990; Fraser and Swinney, 1986; Abarbanel et al., 1990). Here we use an effcient method to estimate the minimum embedding dimension for a correct
reconstruction of the attractor, as introduced by Kennel and Abarbanel (1994), based on the method of false neighbors. The
main idea is to eliminate "illegal projection", finding, for each embedding vector, the nearest neighbor in different
embedding dimensions. When the fraction of false nearest neighbor is less than some threshold, we are able to find the minimum embedding dimension.
The choice of delay time $\tau$ is here based on the so called mutual information of Fraser and Swinney, 1986, which is more adequate than a linear autocorrelation function, when nonlinear dependences are present. The optimal evaluation of this parameter is very crucial
to correctly reconstruct the embedding vectors and their dynamics. Thus, the reliability of the final prediction is strongly dependent from
the above selected parameters.
After a complete characterization of the basic nonlinear dynamics governing the selected time series, we are able to construct a prediction model based on nonlinear
deterministic behavior of the related embedding vectors. Here we follow essentially the approach proposed
in Farmer (1987) and suggested in Zhang (1996) to determine a smooth map for the inverse problem. More precisely,
the nonlinear deterministic behavior in the embebdding space implies the existence of a smooth map: $f^T$ satisfying the relation:

\begin{equation}
 f^T \textbf{$f(y_t)$} = \textbf{$f(y_{t+T})$}
  \end{equation}

for a given embedding vector: $f(y)$. The inverse problem consists on the computation of this smooth map, given a scalar
time series: $x(t), t = 1,2, ..., n$. This map is the basis of our prediction model. Following the approach given in Zhang
(1996) we first divided the known time series into two parts; the first one: $x(t), t = 1, ..., n_0$ is used to set up the
smooth map $f^T$ , and the other part: $x(t), t = n_0+1, ..., n$ is used to check the accuracy of the prediction model. 
In order to calculate the unknown smooth function in the above equation, we assume a local linear hypothesis for the evolution of
the embedding vectors, and this is quite safe for small T values.
Given the last known embedding vector, we selected the first k neighboring vectors near the reference vector in the m-dimensional space,
using a distance function. Then, we assumed that the evolution of the selected vector is correlated with the evolution of the neighboring vectors and
the parameters of this correlation are computed through the solution of a proper least squares problem in the embedding space.
For more mathematical details on this nonlinear approach and its performances to predict solar cycle activity see Sello, 2001.

\section{Results}

In the following, we describe the main results obtained using the above nonlinear dynamics approach when applied to long-term reconstructed TSI time series.
The long-term reconstruction of TSI is essential to apply a nonlinear prediction method which is based on a well known long-term behavior of the selected proxy, here necessary to reliably reconstruct 
an embedding vector dynamics. 

\subsection{Embedding parameters}

The two essential embedding parameters are the delay time, $\tau$ for selecting the vector components, and the embedding dimension, $m$ i.e. the
space to represent the vector dynamics. The computation of the mutual information on the monthly averaged reconstructed TSI values, is shown in Figure 5.
The suggested optimal value for the delay time is the first minimum of the mutual information, here near to: $50 t_s$.

  \begin{figure}[h!]
\resizebox{\hsize}{!}{\includegraphics{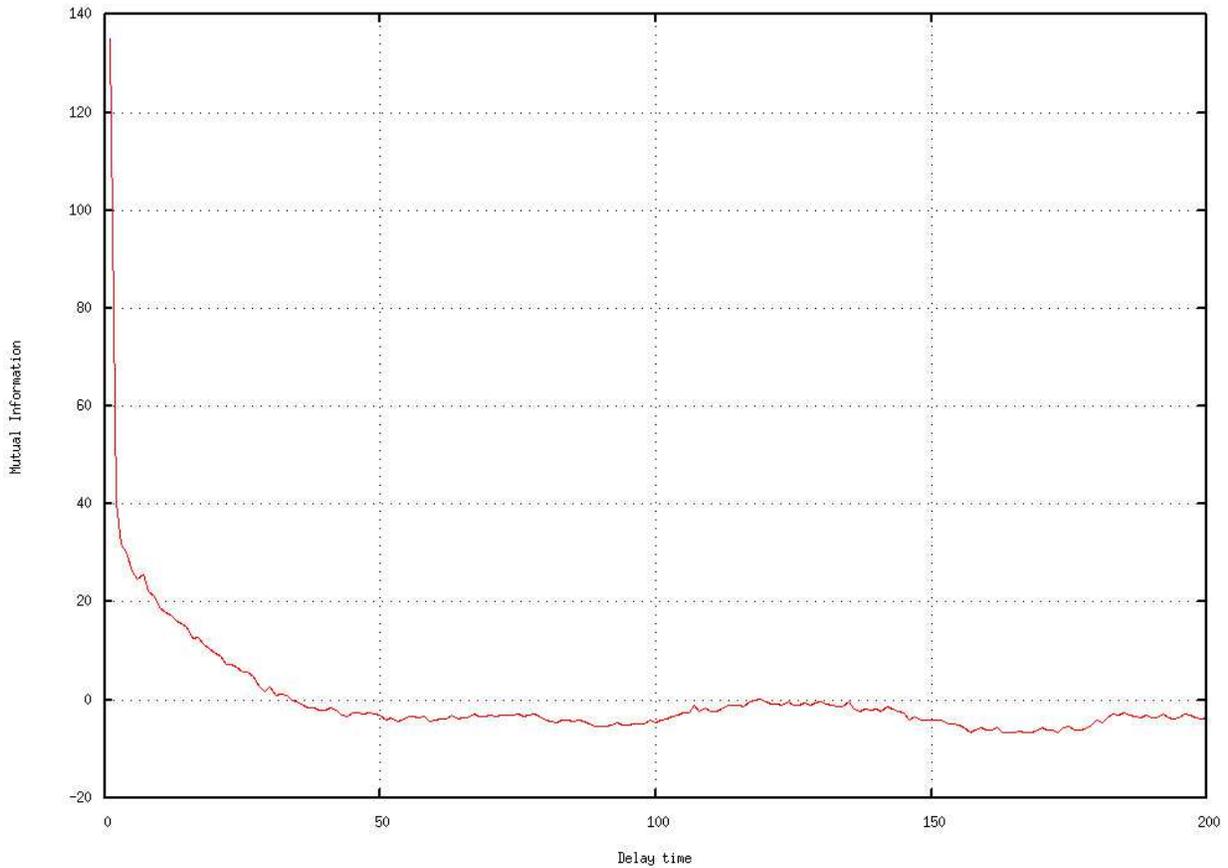}}
 \caption{The computed mutual information for the monthly averaged reconstructed TSI. The optimal delay time is related to the first minimum.}
 \label{fig5}
\end{figure}

The embedding dimension for the reconstructed TSI is evaluated using the method of false neighbors. Figure 6 shows the result: here the
minimum embedding dimension required is: $m=6$.

\begin{figure}[h!]
\resizebox{\hsize}{!}{\includegraphics{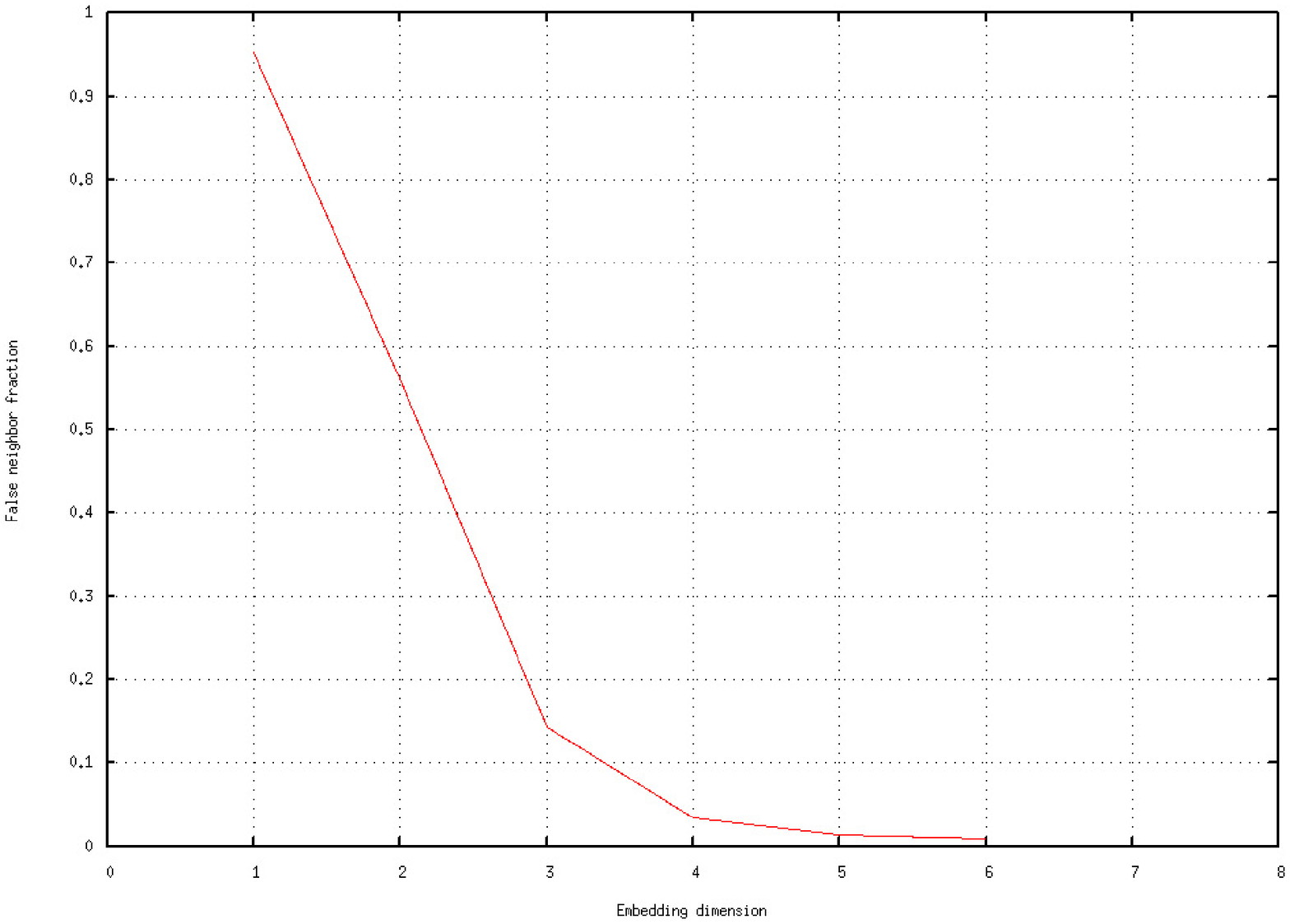}}
 \caption{The embedding dimension derived from the method of false neighbors.}
 \label{fig6}
\end{figure}

The above estimated embedding parameters allow now the correct reconstruction of the nonlinear dynamics for TSI and the application of the
nonlinear prediction method.

\subsection{TSI predictions}

The whole procedure of the nonlinear prediction method has been applied iteratively in order to predict a significant part of the future 
activity. The initial input data, for the monthly averaged TSI, was 4510 and we needed two complete computations to predict a full cycle of
about 11 years. Here we reliably predicted three cycles ahead (six complete computations) starting from 2005.
Figure 7 shows the predicted behavior of TSI for the monthly smoothed values (blue line).
A comparison with the known monthly smoothed TSI values (green line), shows that the three next cycles of solar activity
tend to a clear significant reduction. The most surprising aspect of this prediction is that by knowing only the data of TSI
up to the beginning of 2005, the method has provided the correct phase of the current cycle 24, i.e. the unusual prolonged minimum phase of cycle 23,
followed by a peak (maximum) of cycle 24 predicted for the end of 2012 or early 2013. At that time, none of the prediction methods available, both precursor and non-precursor,
has been able to reliable predict both the observed prolonged minimum of cycle 23 and the current predicted timing of the next maximum.
Further, using the reconstructed TSI data, the method is able to perform statistically reliable long-term predictions, for at least three cycles ahead.

\begin{figure}[h!]
\resizebox{\hsize}{!}{\includegraphics{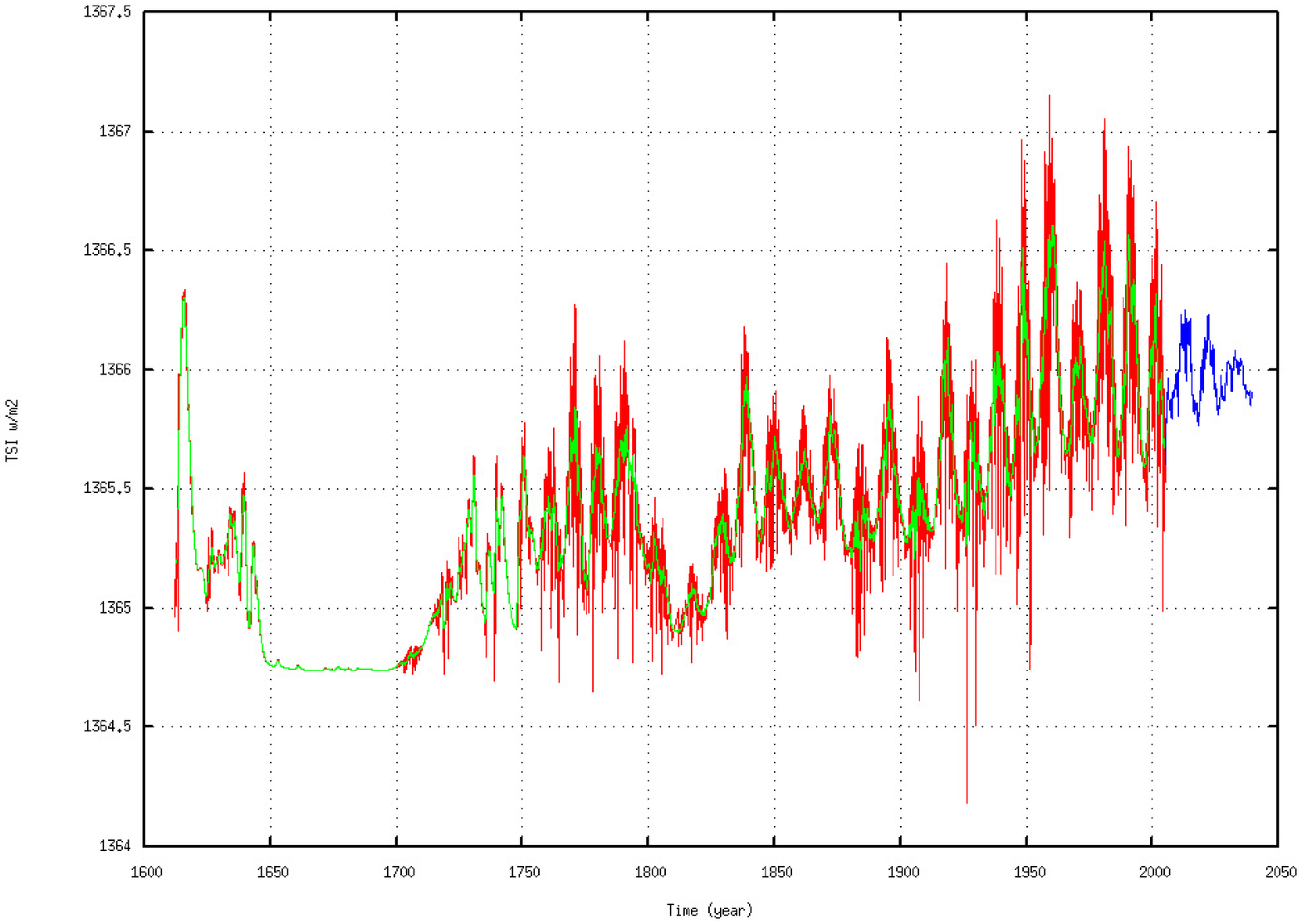}}
 \caption{The predicted monthly smoothed TSI using a nonlinear method starting from 2005 (blue line). 
A comparison with the known monthly smoothed TSI values (green line), shows a clear reduction of activity for the next three cycles.}
 \label{fig7}
\end{figure}

More precisely, the peak predictions based on reconstructed TSI data are: cycle 24 - peak $1366.24 ~ W/m^2$, timing near 2012.7; cycle 25 - peak $1366.21 ~ W/m^2$, timing near 2021.8; cycle 26 - peak $1366.08 ~ W/m^2$,
timing near 2032.6. To better evaluate the shapes of the predicted cycles and their time extensions, Figure 8 shows a proper enlargement. Note the smaller amplitude of predicted cycles.

 \begin{figure}[h!]
\resizebox{\hsize}{!}{\includegraphics{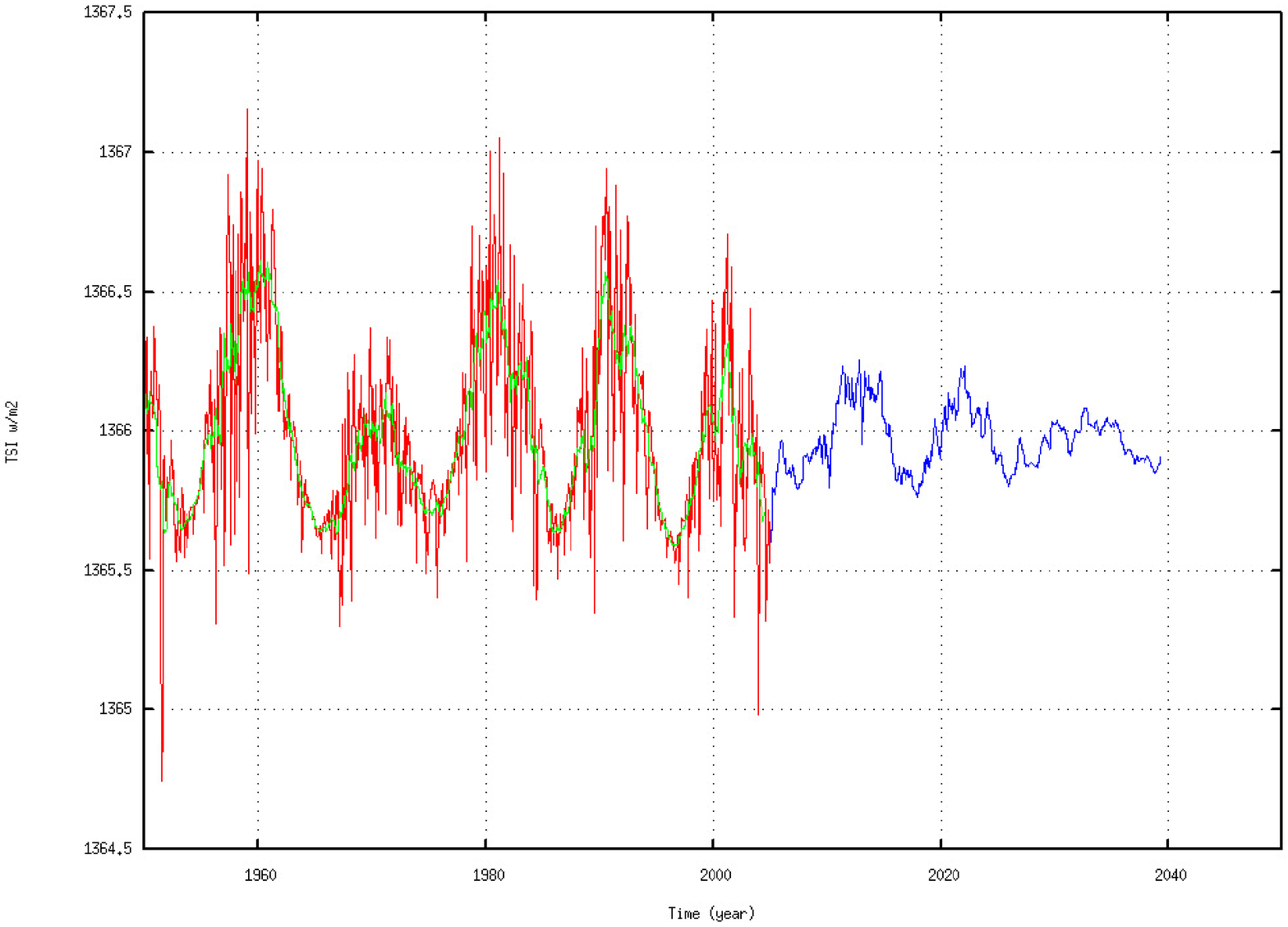}}
 \caption{An enlargement of Figure 7 showing the details of predicted monthly smoothed TSI using a nonlinear method starting from 2005 (blue line).}
 \label{fig8}
\end{figure}

\subsection{F10.7 index and sunspot derived predictions}

From the derived linear correlation between the monthly averaged F10.7 flux peaks and the TSI peaks, we are able
to extrapolate the predicted statistical values for the solar radio flux activity. In the following table we show the related peak values.

\begin{center}
\begin{tabular}{ | l | l | l | }
\hline
Cycle number & Decimal TSI (peak) 1366+ & F10.7 flux (peak) \\ \hline
19 & 0.61 & 223.8 \\ \hline
20 & 0.10 & 140.7 \\ \hline
21 & 0.54 & 184.7 \\ \hline
22 & 0.55 & 191.0 \\ \hline
23 & 0.32 & 176.7 \\ \hline
24 & 0.24 & 156.6 \\ \hline
25 & 0.21 & 153.8 \\ \hline
26 & 0.08 & 136.0 \\ \hline
\end{tabular}
\end{center}

Moreover, using the empirical equation for converting the F10.7 flux to sunspot numbers (Equation (2)), we can easily extrapolate the averaged statistical
values for the maximum monthly smoothed sunspot numbers for the next three cycles. The approximate estimated values are: 99; 97; 81 for cycles 24; 25 and 26,
respectively. It is important to stress again that these predictions are based on data known up to the beginning of 2005. The application
of the nonlinear dynamics method to long-term reconstructed TSI allows a long-term prediction of solar activity and, in particular, appears here more
powerful than other proxies to forecast the phase of future solar cycles, even in the case of quite anomalous behaviors, such as the long prolonged minimum of solar cycle 23,
probably due to the effect of changing or fluctuations in the meridional flow, as suggested by Nandy et al., 2011. In their model, very deep minima are associated with weak polar fields.
The new polar fields, weaker by about $40\%$ when compared to the three previous cycles, were formed several years before the next increase of the meridional circulation, i.e. around mid 2003.
This fact reduces the efficiency of the process for toroidal field amplification and would hence result in a sunspot cycle which is not too strong, with a related weak solar output.
Thus, the TSI index could contain, well in advance, the basic information on the effects due to a change in the amplitude of polar fields responsible, through variations in the meridional circulation, of the observed deep minimum of 
cycle 23. This could explain the ability of the nonlinear method, when using the long-term reconstructed TSI index, to predict well in advance the correct phase of the next cycle, including the unusual deep and long minimum
observed.
Possible connections between the Sun and the Earth's climate are widely documented in literature (see e.g. Hoyt and Schatten, 1997; Haigh et. al., 2004). In particular, the TSI provides the energy that determines the Earth's climate.
As early suggested by Friis-Christensen and Lassen, 1991, the lenght of the solar cycle is closely associated with climate. In particular, the authors
suggest a direct influence of solar activity on global climate through the variation of the solar cycle length. The analyzed data set closely matches the long-term variations of the Northern Hemisphere land air temperature during the past 130 years. 
In 1996, Butler and Johnson demonstrated the same relationship on climate data from the Armagh observatory in Northern Ireland.  As more recently suggested both by Archibald, 2008 and Solheim et al., 2012, the length of the previous sunspot cycle (PSCL) computed between minima has a predictive power for the mean
temperature in the next sunspot cycle. In particular, "the weak or non significant correlations between the temperature and length of the same cycle and the much stronger correlation  for time lags of the order  of 10-12 years, makes the length of the
solar cycle  a good predictor for the average temperature in the next cycle" (Solheim et al., 2012). The authors found that: "...both for the Europe60 stations average and HadCRUT3N acceptable correlations
are found with PSCL and reasonable good predictions are possible". In particular, the authors suggest a linear correlation, $y=\beta x + \alpha$
where $y$ is the temperature or temperature anomaly and $x$ is the time or length  of sunspot cycles. If we consider the HadCRUT3N (Northern emisphere) anomaly temperature
(Brohan et al., 2005), we found: $\beta=-0.207$ and $\alpha=2.11$, with a correlation coefficient: $R= -0.75$. Thus, if we combine the above TSI predictions for the length of a given sunspot cycle and the empirical relation between the PSCL and
the average temperature anomaly in the next cycle, we obtain the results shown in the following table.
    
\begin{center}
\begin{tabular}{ | l | l | l | | l | | l | }
\hline
Cycle number & Start & End & Dur. (yrs) & Temp. An. \\ \hline
23 & 1996.7 & 2008.9 & 12.2 & 0.49  \\ \hline
24 & 2008.9 & 2017.9 & 9.1  & -0.41 \\ \hline
25 & 2017.9 & 2026.1 & 8.2  & 0.23 \\ \hline
26 & 2026.1 & 2038.8 & 12.7 & 0.41 \\ \hline
27 & ---   & ---   & --- & -0.52 \\ \hline
\end{tabular}
\end{center}

\begin{figure}[h!]
\resizebox{\hsize}{!}{\includegraphics{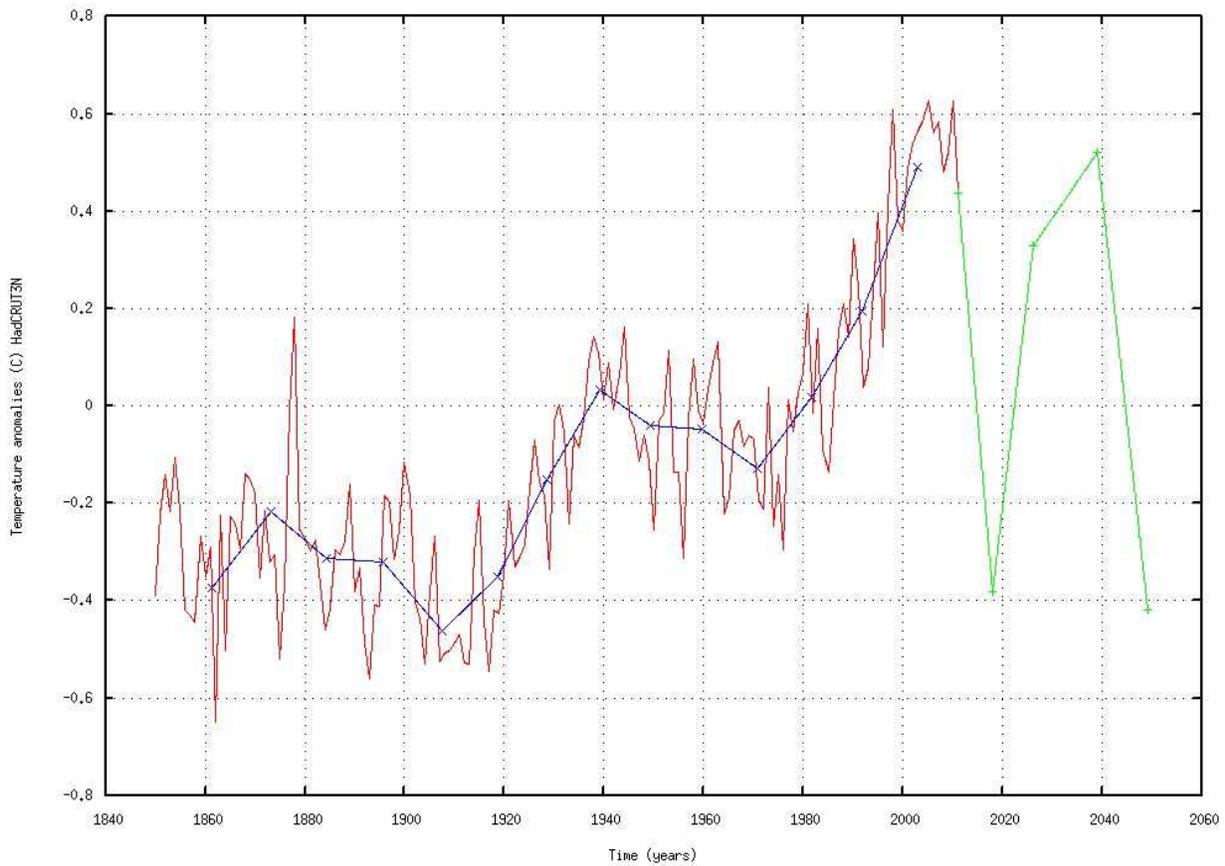}}
 \caption{HadCRUT3N yearly averaged temperature anomalies behavior (red line); the cycle (10-12 years) averaged (blue line) and predicted values for the next four cycles (green line) using both the TSI predictions and the
suggested linear correlation between solar cycle length and the average temperature in the next cycle.}
 \label{fig9}
\end{figure}

Figure 9 shows the related behavior for predicted HadCRUT3N cycle averaged temperature anomalies.
The above result appears coherent with some previous suggestions on the future trend of average temperature anomalies based on solar activity. 
de Jager and Duhau (2011) conclude that the solar activity is presently going through a short transition period (2000-2014), which will be followed
by a Grand Minimum of the Maunder type, most probably starting in the twenties of the current century. Another prediction, based on reduced solar
irradiance due to reduced solar radius, is a sequence of quite lower solar activity cycles leading to a Maunder like minimum starting around 2040 (Abdussamatov,
2007). It is well known that the Maunder Minimum in sunspot numbers in the second half of the seventeenth century coincided with what has become known as the "Little Ice Age "
during which western Europe experienced significantly cooler temperatures. Here we found that, after a significant reduction of the temperature anomalies during the current cycle 24, we could have a pause during the following cycles 25 and 26
with a new average temperature rise (large fluctuations), followed by a significant strong downward of temperature anomalies around 2039-2040 and during the cycle 27.
Of course, the above predicted temperature anomalies mainly consider the solar contribution (R=-0.75) and therefore it would be necessary to include other contributions (systematic errors),
both natural and anthropogenic, added to the statistical errors, to have a more reliable estimate of the actual future temperature anomaly values.
To this regard, it will be particularly interesting to estimate the prediction errors of the average temperature anomalies for the solar cycle 24 as these are not affected by the error
on the estimation of the length of the previous cycle, and then they will provide a lower limit to the overall uncertainty of the prediction method due both to the known statistical
and systematic errors (known unknowns) and to the errors due to different kind (both natural solar or extrasolar and anthropogenic) of unknown components (unknown unknowns).

\section{Conclusions}

The long-term reconstruction of the total solar irradiance since 1611 to the present, based on variations of the
surface distribution of the solar magnetic field and group sunspot numbers by Krivova et al. 2007, allow the reliable use of the nonlinear dynamics method to predict, well in advance,
the solar activity for at least three cycles ahead. From the derived linear correlation between the monthly averaged F10.7 flux peaks and the TSI peaks, we are able
to extrapolate the predicted statistical values for the solar radio flux activity. Further, using the empirical equation for converting the F10.7 flux to sunspot numbers,
we can easily extrapolate the averaged statistical values for the maximum monthly smoothed sunspot numbers for the next three cycles. 
The above analysis supports other recent independent analyses, suggesting and confirming a general reduction trend for the future solar activity.
Taking into account the link existing between the total solar output and the earth's climate response, the ability to perform long-term
predictions of solar activity many years in advance, appears as a very interesting tool in order to derive useful information not only 
for specific solar activity studies and for space technology activities, but also for better modelling the future global climate evolution.

\section{References}

Abarbanel, H. D. I., Brown, R.,  Kadtke, J. B. 1990: Prediction in chaotic nonlinear systems - Methods for time series with broadband Fourier spectra,
Phys. Rev. A, 41, 4.

Abdusamatov, K.I. 2007: Optimal Prediction of the Peaks of Several Succeding Cycles on the Basis of Long-Term Variations in the Solar Radius
or Solar Constant, Kinematics and Physics of Celestial Bodies, 23,9 7-100.

Archibald D. 2008 : Solar cycle 24, Rhaetian Management Pty Ltd. 29 Pindari Road, City Beach, WA 6015, Australia

Brohan, P., Kennedy, J.J., Harris, I., Tett S.F.B., and Jones, P.D. 2005: Uncertainty estimates in regional and global observed
temperature changes: a new dataset from 1850, Jour. Geophys. Res., 111, D12106.

Butler, C. J., Johnston, D. J., 1996: A provisional long mean air temperature series for Armagh Observatory Authors, Jour. Atm. and Terrestrial Physics, 58, 1657-1672.

de Jager, C. and Duhau, S., 2009: Forecasting the parameters of sunspot cycle 24 and beyond, Jour. Atm. Solar-Terr. Phys., 71, 239.

de Jager, C. and Duhau, S. 2011: The variable solar dynamo and the forecast of solar activity; influence on terrestrial surface temperature, in Global
Warming in the 21th Century, J. M. Cossia (ed), Nova Science Publ. N.Y., pp. 77-106.

Farmer, J. D., and Sidorowich, J. J. 1987: Predicting chaotic time series, Phys. Rev. Lett., 59,8.

Fraser, A. M., and Swinney, H. L. 1986:  Independent coordinates for strange attractors from mutual information, Phys. Rev. A, 33, 2.

Friis-Christensen, E., Lassen, K. 1991: Length of the Solar Cycle: An Indicator of Solar Activity Closely Associated with Climate, Science, 254, 5032. 

Haigh, J.D., Lockwood, M., Giampapa, M.S. 2004: The Sun, Solar Analogs and Climate, Saas Adavanced Course 34, Springer Berlin.

Hoyt, D.V., Schatten, K.H., 1997: The Role of the Sun in Climate Change, Oxford University Press.

Kennel, M. B., and Abarbanel, H. D. I. 1994:  Prediction Errors and Local Lyapunov Exponents, Phys. Rev. E, 47,4.

Krivova, N.A., Balmaceda L., and Solanki S. K., 2007: Reconstruction of solar total irradiance since 1700 from the surface magnetic flux, A\&A 467, 335–346.

Kugiumtzis, D. 1999, Los Alamos National Laboratory Preprint Archive, submitted to Phys. Rev. E [Physics/9905021].

Lawrence, J. K., Cadavid, A. C., Ruzmaikin, A. A., 1995:  Turbulent and Chaotic Dynamics Underlying Solar Magnetic Variability, ApJ, 455.

Li, L. H., and Sofia, S. 2001: Measurements of Solar Irradiance and Effective Temperature as a Probe of Solar Interior Magnetic Fields, ApJ, 549, 2, 1204.

Lockwood, M. 2005, in The Sun, Solar Analogs and the Climate, 34th Saas Fee Advanced Course, ed. I. Rüedi, M. Güdel, and W. Schmutz (Berlin: Springer),
109.

Nandy, D., Munoz-Jaramillo, A., Martens, P. C. H., 2011:  The unusual minimum of sunspot cycle 23 caused by meridional plasma flow variations,
Nature, 471, 80.

Palus, M. 1995:  Detecting nonlinearity in multivariate time series, Physica D, 80, 186.

Pesnell, W. D., NOAA/SEC prediction panel www.sec.noaa.gov/SolarCycle/SC24/index.html.

Prichard, D., and Theiler, J. 1994:  Generalized redundancies for time series analysis, Phys. Rev. Lett., 73, 951.

Rozelot, J. P. 1995: On the chaotic behaviour of the solar activity,  A\&A, 297, L45.

Schreiber, T. 1998:  Interdisciplinary application of nonlinear time series methods, Phys. Rep., 308.

Sello, S., 2012: A non-linear full-shape curve prediction after the onset of the new solar cycle 24, Jour. Atm. Solar-Terr. Phys.,
Vol. 80, 15 May 2012, Pages 252-257.

Sello, S., 2001: Solar cycle forecasting: A nonlinear dynamics approach, A\&A, 377, 1, 312.

Solheim, J.E., Stordahl, K., Humlum, O., 2012, The long sunspot cycle 23 predicts a significant temperature decrease in cycle 24, Jour. Atm. Solar-Terr. Phys.,
In Press, Corrected Proof, Available online 16 February 2012.

Solanki, S. K., Krivova, N. A. 2004: Solar Irradiance Variations: From Current Measurements to Long-Term Estimates, Sol. Phys., 224, 197.

Solanki, S. K., Schüssler, M., Fligge, M. 2000: Evolution of the Sun's large-scale magnetic field since the Maunder minimum,  Nature, 408, 445.

Solanki, S. K., Schüssler, M., Fligge, M. 2002: Secular variation of the Sun's magnetic flux, A\&A, 383, 706.

Theiler, J. 1990: Estimating fractal dimension, Jour. Opt. Soc. Am. A, 7, 6.

Wenzler, T., Solanki, S. K., Krivova, N. A., Fröhlich, C. 2006: Reconstruction of solar irradiance variations in cycles 21-23 based on surface magnetic fields, A\&A, 460, 583.

Zhang, Q. 1994:  Research on Fractal Dimension for Sunspot Relative Number, Acta Astron. Sin., 35.

Zhang, Q. 1995: Predictability of the long term variations of monthly mean sunspot numbers, Acta Astron. Sin., 15.

Zhang, Q. 1996: A nonlinear prediction of the smoothed monthly sunspot numbers, A\&A, 310, 646.

\end{document}